\documentclass[prl,twocolumn,superscriptaddress]{revtex4}

\usepackage{epsfig}

\input{Qcircuit.tex}

\newcommand{\kb}[2]{ | #1 \rangle \langle #2 |}
\newcommand{\tr}{\mathrm{tr}}
\newcommand{\union}{\cup}
\newcommand{\order}{O}
\newcommand{\poly}{\mathrm{poly}}

\begin{document}

\title{Efficient classical simulation of the approximate quantum Fourier transform}

\author{Nadav Yoran}\email{N.Yoran@bristol.ac.uk} \author{Anthony J. Short}
\affiliation{H.H.Wills Physics Laboratory, University of Bristol,
Tyndall Avenue, Bristol BS8 1TL, UK}

\begin{abstract}
 We present a method for classically simulating quantum circuits
 based on the tensor contraction model of Markov and Shi (quant-ph/0511069).
 Using this method we are able to classically simulate the approximate quantum Fourier
 transform in polynomial time. Moreover, our approach allows us to formulate a
 condition for the composability of simulable quantum circuits. We
 use this condition to show that any circuit composed of a constant
 number of approximate quantum Fourier transform circuits and $\log$ depth circuits
 with limited interaction range can also be efficiently simulated.

\end{abstract}

\maketitle

\noindent One of the most useful ways of investigating the power,
and limitations, of quantum computation is to identify classes of
quantum algorithms which can be efficiently simulated on a classical
computer. A well-known example of such a class are circuits composed
of Clifford group operations, which are shown by the Gottesman-Knill
theorem \cite{gottesman} to be efficiently simulable. Recently a
number of new methods for simulating quantum computation have
appeared \cite{jozsa, markov, briegel, us}, both for the circuit
model and the measurement-based model of quantum computation. Unlike
the Gottesman-Knill theorem which is based on restricting the set of
allowed gates, all these new methods rely on the topology (the
`graph' of connections) of the simulated circuit. In particular two
of these new methods, due to Jozsa \cite{jozsa} and Markov and Shi
\cite{markov}, both use the formalism of tensor contraction for
simulations in the quantum circuit model, which is the focus of this
paper.

We base our approach on Markov and Shi's formalism, which has the
advantage of being able to simulate generalised quantum dynamics and
mixed states (this would be particularly useful in simulating noisy
gates), as well as working directly with the natural graph of the
circuit. Using this approach, we show how the approximate quantum
Fourier transform (AQFT) can be efficiently simulated on a classical
computer (i.e. simulated in a time polynomial in the number of input
qubits). Additionally, our method allows us to formulate a simple
condition for the composability of two simulable circuits. That is,
if the simulation procedures for two circuits obey a particular
condition, we are assured that the composed circuit (created by
connecting the outputs of one to the inputs of the other) will also
be efficiently simulable. We use this condition to show that any
circuit composed of constant number of AQFT circuits and $\log$
depth quantum circuits with bounded interaction range can be
efficiently simulated on a classical computer. Obviously, this
implies that the AQFT can be efficiently simulated when applied to
any state produced by such circuits.

\textbf{Simulating Quantum Computation} In order to simulate a
quantum computation, we first associate a graph with the circuit in
the obvious way, representing each input qubit, gate, and output
qubit by a vertex, and each wire by an edge (e.g. a two-qubit gate
would correspond to a vertex of degree four). Next, we label each
edge with a different index (i,j,k, etc.). Each index ranges over
four possible values, corresponding to the four components of a
qubit's density operator. Finally, to each vertex we associate a
tensor describing the operation performed at that point. This tensor
has indices corresponding to all edges connected to that vertex (so
that its rank is equal to the degree of the vertex). For clarity, we
use raised indices to denote output wires, and lowered indices to
denote input wires.

Following Markov and Shi's approach \cite{markov}, we associate
tensors with basic circuit elements as follows, using the operator
basis $e_i = \{\kb{0}{0}, \kb{0}{1}, \kb{1}{0}, \kb{1}{1} \}$ for
single qubits, and $e_{ij}=e_i \otimes e_j$ for two qubits:

\begin{enumerate}

\item Inputting a qubit in state $\rho$:
\begin{equation}
\Qcircuit @C=0.5cm @R=1cm {  & \lstick{\rho}  & \ustick{i} \qw & \qw
}  \qquad T^i =  \tr(e^{\dagger}_i \rho)
\end{equation}

\item Performing a single-qubit operation $\rho \rightarrow G[\rho]$:
\begin{equation}
\Qcircuit @C=0.5cm @R=1cm {   & \ustick{i} \qw & \gate{G} &
\ustick{j} \qw & \qw } \qquad T^{j}_{i} = \tr(e^{\dagger}_{i}
G[e_{j}])
\end{equation}

\item Performing a two-qubit operation $\rho \rightarrow G'[\rho]$
\begin{equation}
\Qcircuit @C=0.5cm @R=0.6cm {   & \ustick{i} \qw & \multigate{1}{G'}
& \ustick{k} \qw & \qw \\ & \ustick{j} \qw & \ghost{G'} & \ustick{l}
\qw & \qw } \qquad T^{k l}_{i j} = \tr ( e^{\dagger}_{ij}  G'[
e_{kl} ])
\end{equation}

\item Obtaining a measurement result corresponding to a generalised measurement (POVM) operator $E$:
\begin{equation}
\Qcircuit @C=0.5cm @R=1cm {   & \ustick{i} \qw & \measuretab{E} }
\qquad T_{j} = \tr(E e_j)
\end{equation}

\item Discarding a qubit, or obtaining an unspecified measurement result:
\begin{equation}
\Qcircuit @C=0.5cm @R=1cm {   & \ustick{i} \qw & \measuretab{}  }
\qquad T_{j} = \tr(e_j)
\end{equation}

\end{enumerate}

Note that these examples can easily be extended to apply to joint
input states or measurements, gates acting on more qubits, or gates
with different numbers of inputs and outputs. Tensors could even be
introduced to represent non-physical (i.e. not completely positive)
linear operations if desired.

\begin{figure}\begin{center}

\epsfig{file=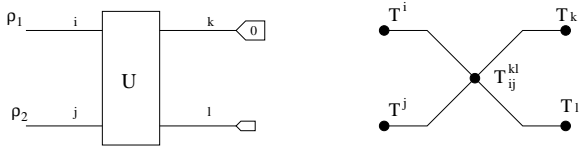} \caption{The graph and tensors associated
with a simple example circuit described below, in which a two-qubit
unitary gate acts on the state $\rho_1 \otimes \rho_2$, and the
first qubit is measured to be in state $\ket{0}$.\label{fig:gate} }
\end{center}\end{figure}

Once a tensor with the appropriate indices has been assigned to each
vertex, taking their product and summing over all indices will yield
the probability of obtaining the specified measurement result. This
process is illustrated below for the simple circuit shown in fig.
\ref{fig:gate}, in which a two qubit unitary gate $G[\rho] = U\rho
U^{\dagger}$ acts on the separable input state $\rho_1 \otimes
\rho_2$, and the probability $p$ of the first qubit being found in
the state $\ket{0}$ is obtained.
\begin{eqnarray}
p &=&  \sum_{ijkl} T^i\, T^j\, T^{kl}_{ij}\, T_k\, T_l \label{eqn:gate1} \\
&=& \sum_{ijkl} \tr(e^{\dagger}_i \rho_1)\tr(e^{\dagger}_j
\rho_2)\tr( e^{\dagger}_{kl} U e_{ij}
U^{\dagger}) \tr(\kb{0}{0} e_k)\tr(e_l)  \nonumber \\
&=& \sum_{kl} \tr( e^{\dagger}_{kl} U (\rho_1 \otimes \rho_2)
U^{\dagger})
\tr( (\kb{0}{0} \otimes I) e_{kl}) \nonumber \\
&=&  \tr \left( (\kb{0}{0} \otimes I) U (\rho_1 \otimes \rho_2)
U^{\dagger} \right) \nonumber
\end{eqnarray}

Note that in cases where we wish to measure many output qubits, it
may be prohibitive to calculate the probability of all possible
output strings (as there are exponentially many possibilities).
Instead, a closer analogy to the real quantum computation would be
to sample from the probability distribution to obtain a particular
random measurement result. This can be achieved by computing the
probability of measurement outcomes for the first qubit (with the
measurement results for all other qubits unspecified) and randomly
selecting a result, then computing the joint probability of
measurement results for the second qubit with the chosen result on
the first qubit, and randomly selecting one, and so on until a
particular measurement result has been selected for each qubit.

The problem with summing over all tensor indices at the same time
(as written in equation (\ref{eqn:gate1})) is that there are
exponentially many terms, making the computation very slow. To avoid
this, we `contract' the tensors together one at a time, breaking the
joint sum into a series of separate sums. In each step of the
computation we replace two existing tensors with a new tensor
obtained by summing over any repeated indices (e.g. $T^{kl}_{ij}
T^{no}_{lm} \rightarrow T^{kno}_{ijm}$). We repeat this procedure
until we are left with a single tensor with no free indices, which
is the desired probability. The aim is to order the contractions so
that we never generate tensors with too many indices during this
process.

In Markov and Shi's paper, they describe this contraction process by
an ordering on the edges of the graph (i.e. on index summations).
However here we take a different approach, in which the contraction
process is described by a sequence of sets of vertices $\cal{S}$
$=(s^{1},\ldots,s^{N})$ - each of which corresponds to a particular
tensor that is generated during the computation. This allows us to
formulate a condition for efficient simulation of composite
circuits.

The tensor corresponding to a set of vertices $s$ is that generated
by contracting together all initial tensors corresponding to
vertices in $s$. In each step of the contraction process we take two
existing tensors and generate a new one, so each set $s^i \in
\cal{S}$ is either the union of two previous sets, or one previous
set and a vertex, or two vertices. Denoting the set of all vertices
by $V$:

\begin{equation}
 s^{i}=\{t_{1}^{i}\cup t_{2}^{i}\} \quad \mbox{where} \quad \begin{array}{l}
 \mbox{either}\;\;\: t_{j}^{i}=s^{k}\: , \: k<i \: , \\ \mbox{or}\quad\quad \:
 t_{j}^{i}=\{v\} \: , \:v \in V. \end{array}
 \end{equation}

The calculation of the probability is done in $N$ steps, where in
step $i$ we compute a new tensor by summing over all indices
corresponding to edges connecting $t_{1}^{i}$ to $t_{2}^{i}$. For
the computation to be complete, we require that the final set
$s^{N}=V$. Note that sampling from the output probability
distribution for many qubits as described above only requires
changing the measurement operators applied to the outputs, and hence
each run can use the same graph and contraction sequence ${\cal S}$.

The computational difficulty of the simulation is determined by the
maximal rank of the tensors generated during the computation. For
each $s^{i}$ in $\cal{S}$ we therefore define $E^{i}$ as the number
of edges that connect vertices in $s^{i}$ to vertices outside
$s^{i}$, which is exactly the rank of the tensor corresponding to
$s^{i}$. The simulation corresponding to the sequence $\cal S$ will
be an efficient one if $E^{max}=\max_i E^{i}=\order(\log n)$. This
condition assures us that the maximal number of components for each
tensor we compute is $\order(\poly(n))$. Furthermore, the maximal
number of terms summed over when computing each component must also
be $\order(\poly(n))$, as the two sets  $t_{1}^{i}$ and $t_{2}^{i}$
which are combined to form $s^i$ can only connect on at most
$E^{max}=\order(\log n)$ edges.

A class of circuits that is easy to simulate efficiently using this
approach is that of log-depth circuits with bounded interaction
range (i.e. involving $d=\order(\log n)$ timesteps, in which gates
act on qubits at most a constant distance $r$ apart).  To simulate
such circuits, we number all vertices involving qubit 1 first (i.e.
gates, inputs and outputs on the upper horizontal line of the
circuit), then all vertices involving qubit 2 that are not already
included, and so on until we have numbered all the vertices. The
sequence $\mathcal{S}$ is then composed of sets containing
increasing numbers of vertices in this ordering (e.g. $(\{v^1\},
\{v^1, v^2\}, \ldots )$. It is easy to see that for such
simulations, $E^{\max} \leq d r = \order(\log n)$. The efficient
simulability of such circuits has previously been shown in
\cite{markov, us}. Furthermore, it was shown by  Jozsa \cite{jozsa},
using a slightly different approach \footnote{Note that Jozsa's
construction can be represented in our formalism by `folding' the
inverse circuit on top of the standard circuit and combining the
tensors and indices vertically.}, that the same strategy will work
for any circuit in which each qubit line is touched (or crossed) by
at most $\order(\log n)$ gates.

\smallskip
\textbf{Simulating the Approximate Quantum Fourier Transform}. An
important new example of a circuit which can be efficiently
simulated by the above scheme is the approximate quantum Fourier
transform (AQFT). An efficient circuit for the exact Fourier
transform \cite{nielsen} consists of a sequence of $(n-1)$ `ladder
circuits' of decreasing size.  The $l^{\mathrm{th}}$ ladder circuit
is composed of a Hadamard gate on the $l^{\mathrm{th}}$ qubit,
followed by ($n-l$) conditional phase gates connecting qubit $l$ to
qubits $l+1,\ldots n$ respectively. These conditional phase gates
have the form $R_k=exp(\pi i/2^k) \kb{11}{11}$, where $k$ is the
distance over which the gate acts.
\begin{figure}\begin{center}
\epsfig{file=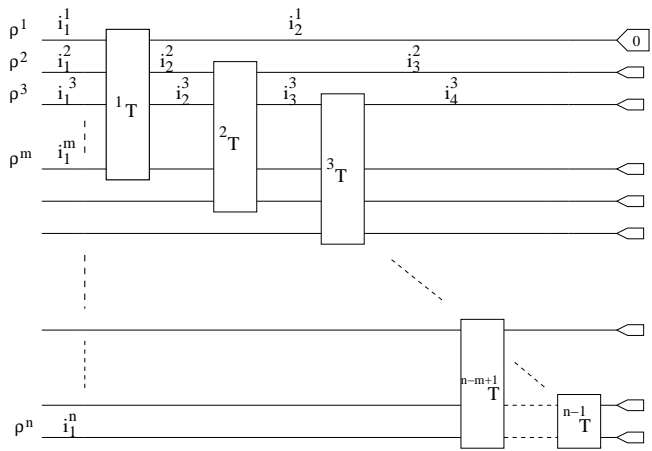} \caption{The general structure of the
circuit calculating the approximate quantum Fourier transform,
showing a measurement of $\ket{0}$ on the first output qubit. Each
box corresponds to a ladder circuit, with internal gates as shown in
fig.\ref{fig:ladder}. \label{fig:aqft} }
\end{center}\end{figure}
However, it was noted by Coppersmith \cite{coppersmith} that in many
case an exact Fourier transform is not necessary, and that a very
good approximation can be obtained by omitting all gates $R_k$ with
$k>m$ (i.e. gates that act over a large distance, and generate only
small phase rotations). In what follows, we take
$m=\log(n/\epsilon)$, yielding an error in the final state of
$\order(\epsilon)$. Furthermore this approximate quantum fourier
transform is sufficient for the most useful application of the
algorithm - for estimating periodicity, and hence for use in Shor's
factoring algorithm \cite{shor}. Barenco \emph{et al.}
\cite{barenco} proved that the AQFT will yield the same probability
of success as the exact periodicity-finding algorithm after
$\order(n^3/m^3)$ runs. A diagram of the AQFT circuit is given in
fig. \ref{fig:aqft}. Note that in this circuit, the output qubits
occur in reverse order to the inputs (i.e. starting at the bottom).
\begin{figure}\begin{center}
\epsfig{file=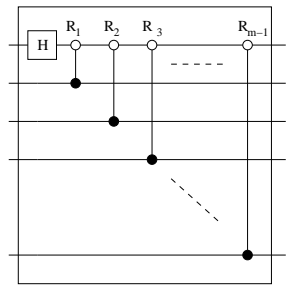} \caption{ The gates composing a ladder
circuit, consisting of a Hadamard gate and then $m-1$ controlled
rotation gates. Note that the last few ladder circuits are actually
slightly smaller, although they have the same form.
\label{fig:ladder} }
\end{center}\end{figure}

In order to classically simulate the $AQFT$ circuit we cannot use
the same simple ordering $\mathcal{S}$ used for the log-depth
circuits above, as $O((\log n)^2)$ gates cross each qubit line. This
leads to tensors with $n^{O(\log\,n)}$ elements, that cannot be
computed in polynomial time. Instead, we choose the following
contraction ordering: We first contract together all of the tensors
corresponding to gates in the first ladder circuit (in any order),
then we proceed to do the same for the second ladder circuit, and so
on, until we have one combined tensor for each ladder circuit (i.e.
until $\mathcal{S}$ contains  sets corresponding to the vertices in
each ladder circuit). Since there are at most $m = \order(\log n)$
two-qubit gates in each ladder circuit, all the tensors we generate
 have at most $\order(\log n)$ indices.

Next we combine the ladder circuits and their associated inputs and
outputs, one by one in descending order, until we have contracted
together all the remaining tensors. First, we take the tensor for
the top ladder circuit and contract it with all the tensors of input
and output vertices to which it is connected, in order from top to
bottom (i.e. $i_{1}^{1}$ to $i_{1}^{m}$ and $i_{2}^{1}$) . Then we
contract this new tensor with the tensor for the second ladder
circuit, and again contract it with any input or output vertices to
which it is connected (i.e. $i_{1}^{m+1}$ and $i_{2}^{2}$). We
continue to contract each new tensor with that of the next ladder
circuit, and its inputs and outputs vertices, until all tensors have
been included - and thus we have computed the probability of the
chosen measurement result.

Note that in each stage of the contraction process, the new tensors
we generate only have at most $E^{max}=\order(\log n)$ free indices
- hence storing and computing these tensors requires only polynomial
time and memory space. We have therefore proved that the approximate
quantum fourier transform can be efficiently simulated classically.

So far we have assumed that the input to the AQFT circuit is a
product state of $n$ qubits (although this need not be a
computational basis state). A way to generalize this set of possible
input states  is to identify classes of efficiently simulable
circuits that can be connected to the inputs of the AQFT circuit
such that the composed circuit is also efficiently simulable. In the
next section, we therefore consider the simulability of composed
circuits.

\smallskip
\textbf{Simulating composed circuits}. Consider two efficiently
simulable circuits, $A$ and $B$, that we join together to form a
composed circuit $C$ by connecting the output wires of $A$ to the
input wires of $B$ (for simplicity we assume that both are $n$ qubit
circuits, our discussion can be easily generalized to the case where
only some of the inputs and output connect). How can we tell if the
composed circuit $C$ is efficiently simulable?

In what follows we will use the subscripts $a$ and $b$ to label
objects belonging to circuits $A$ and $B$ respectively. Let us also
denote the set of output vertices of $A$ by $V_{a}^{out}$ and  the
set of input vertices of $B$ by $V_{b}^{in}$. Given efficient
simulations of $A$ and $B$,  $C$ will be efficiently simulable if
the following condition holds:  For any subset $\omega_a=\{v_{a,
i_1}^{out}\ldots v_{a, i_k}^{out} \}$ of $V_{a}^{out}$, if there is
a set $s_a \in {\cal S}_{a}$ which includes exactly this subset of
output vertices (i.e. $s_a \cap V_{a}^{out}= \omega_a$), then there
is a set $s_b \in {\cal S}_{b}$ containing exactly the same subset
of input vertices (i.e. $s_b \cap V_{b}^{in} = \{v_{b,
i_{1}}^{in}\ldots v_{b, i_{k}}^{in} \}$).


To prove that this condition is sufficient, we first decompose all
the sets in ${\cal S}_{a}$ containing output vertices into their
output and non-output components, writing $s^{ij}_a = \omega^i_a
\union \mu^{ij}_a $, where  $\omega^i_a \in V_a^{out}$ denotes a
particular set of output vertices, and $\mu^{ij}_a$ denotes a
corresponding set of non-output vertices. Similarly, we decompose
each set $s_b \in {\cal S}_{b}$ containing input vertices in the
same way into a set of input vertices $\eta^i_b$ and their
associated non-input vertices $\mu^{ij}_b$, such that $s^{ij}_b =
\eta^i_b \union \mu^{ij}_b$.

From our simulation procedure it is clear that any two sets in a
sequence are either disjoint or are such that one includes the
other. It is also clear that the order in which two disjoint sets
are constructed is arbitrary (we can choose which set to construct
first). Therefore, by re-labelling  and re-ordering the sets in
$\mathcal{S}_a$, we can ensure that all sets not containing output
vertices occur first, and that the remaining sets $s^{ij}_a$ occur
in the order of increasing $i$ and then increasing $j$ (e.g. $
s^{11}_a, s^{12}_a, s^{13}_a, s^{21}, \ldots $), and similarly for
$\mathcal{S}_b$ and the input vertices. Furthermore, our
composability condition ensures that we can find sequences of this
form, such that the output vertices in $\omega^i_a$ connect
precisely with the input vertices in $\eta^i_b$.

We  construct a sequence of sets $\mathcal{S}_c$ for the combined
circuit $C$ as follows: Starting with circuit $A$, we first include
all sets from $\mathcal{S}_a$ that do not involve input vertices,
then do the same for $\mathcal{S}_b$. The next set we include is
$\mu^{11}_a \union \mu^{11}_b$, in which the first output from $A$
is contracted with an input of $B$ (yielding the union of two
non-output sets). We proceed to evolve this set in $A$ by including
$\mu^{1j}_a \union \mu^{11}_b$  for $j=2,\ldots j^{\max}_1$. After
this, we shift to evolving circuit $B$, by including sets
$\mu^{1j^{\max}_1}_a \union \mu^{1k}_b$ for $k=1,\ldots k^{\max}_1$.
Then, beginning with $\mu^{21}_a \union \mu^{21}_b$, we repeat the
above procedure  for $i=2, \ldots, i^{max}$, by including
$\mu^{ij}_a \union \mu^{i1}_b$ for $j=1,\ldots j^{\max}_i$, then
$\mu^{ij^{\max}_2}_a \union \mu^{ik}_b$ for $k=1,\ldots j^{\max}_i$,
until all vertices in the combined circuit have been included.

The key point is that at any stage in the above process the sets we
construct are either identical to a set in the original sequences,
or composed of a union of two such sets with some input and output
vertices discarded (i.e those across which the circuit is
connected). It is therefore clear that $ E^{\max}_c \leq
(E_{a}^{\max}+ E_{b}^{\max})$, and hence when both $E_{a}^{\max}$
and $E_{b}^{\max}$ are  $\order(\log n )$, the simulation process
defined by ${\cal S}_{c}$ is an efficient one.

These results can be generalised to apply to any constant number of
efficiently simulable circuits connected in series. In such cases,
the combined circuit will be efficiently simulable when the above
composability condition is satisfied across each circuit boundary.

\smallskip

From the simulation procedures for the AQFT circuit and log-depth
limited range circuits given above, we see that both the input and
output vertices are included sequentially from bottom to top (i.e.
$\omega_i = \{ v_1^{out}, \ldots, v_i^{out} \}$ and $\eta_i =  \{
v_1^{in}, \ldots, v_i^{in} \}$). As each output set from one circuit
corresponds exactly to an input set for the other, these two
circuits obey our composability condition. Furthermore, with
$\omega_i$ and $\eta_i$ defined as above, their simulation sequences
do not need to be rearranged before they are composed. By joining
these two circuits together, we can classically simulate the
 approximate quantum Fourier transform on any input state that can be produced by
log-depth circuit involving limited range interactions.

Because the outputs of the AQFT circuit occur in reverse order,
attaching a circuit afterwards is more tricky, but can be achieved
by flipping the attached circuit vertically. In order to satisfy the
composability condition, tensors in the flipped circuit must still
be contracted from top to bottom, but this can easily be achieved
for both types of circuit considered here (since the original
circuits are also simulable with a bottom to top contraction
ordering). We therefore conclude that any circuit which is composed
of a constant number of AQFT and log-depth limited range circuits
can be simulated efficiently on a classical computer.

\smallskip
\acknowledgments

The authors wish to thank R. Josza, S. Popescu, A. Montanaro, D.
Browne and H. Briegel for fruitful discussions. The work of N.~Y.
was supported by UK EPSRC grant (GR/527405/01), and A.J.S. was
supported by the UK EPSRC's ``QIP IRC'' project.

\smallskip

Note added: After the completion of this work, we became aware of a
very recent paper by Aharonov, Landau and Makowski
(quant-ph/0611156) which appears to simulate an AQFT circuit in
$n^{O(\log \,n)}$ time.

\end{document}